\documentclass[useAMS,usenatbib]{mn2e}

\usepackage{graphicx}
\usepackage{float}
\usepackage{pdfpages}
\usepackage{longtable}
\usepackage{lipsum}
\usepackage{textcomp,gensymb}

\newcommand{\photu}{ph\ cm$^{-2}$ s$^{-1}$ sr$^{-1}$ \AA$^{-1}$}
\newcommand{\galex}{{\it GALEX}}

\title{Diffuse Radiation from the Aquila Rift}

\author[Jyothy et al.]{S. N. Jyothy$^{1}$\thanks{jyothi.sasidharan.nair@gmail.com}, Jayant Murthy$^{2}$\thanks{jmurthy@yahoo.com }, Narayanankutty Karuppath$^{1}$ and N. V. Sujatha$^{3}$\\
$^{1}$Amrita Vishwa Vidyapeetham, Kollam 690525, India\\ 
$^{2}$Indian Institute of Astrophysics, Bengaluru 560034, India\\
$^{3}$St.Xavier's College for Women, Aluva 683101, India}

\begin{document}
\date{Accepted . Received ; in original form }
\pagerange{\pageref{firstpage}--\pageref{lastpage}} \pubyear{2015}
\maketitle
\label{firstpage}

\begin{abstract}
We present an analysis of the diffuse ultraviolet (UV) background in a low latitude region near the Aquila Rift based on observations made by the {\it Galaxy Evolution Explorer} (\galex). The UV background is at a level of about 2000 \photu\ with no correlation with either the Galactic latitude or the 100 micron infrared (IR) emission. Rather, the UV emission falls off with distance from the bright B2 star HIP 88149, which is in the centre of the field. We have used a Monte Carlo model to derive an albedo of 0.6 -- 0.7 in the UV with a phase function asymmetry factor ($g$) of 0.2 -- 0.4. The value for the albedo is dependent on the dust distribution while $g$ is determined by the extent of the halo.
\end{abstract}

\begin{keywords}
ultraviolet: ISM - diffuse radiation - dust
\end{keywords}

\section{Introduction}
The diffuse ultraviolet radiation field is comprised of sources ranging from airglow and zodiacal light \citep{Murthy2014a} to diffuse galactic light \citep{Hyakawa1969,lillie1976} with a possible extragalactic contribution at high Galactic latitudes \citep{Henry2002}. Early observations in the field and their conclusions were reviewed by \citet{Bowyer1991} and \citet{Henry1991} with a more recent review by \citet{Murthy2009}.

There have been relatively few observations of the diffuse UV light at low latitudes, partly because many of the early observations were looking for an extragalactic background \citep{sasseen1995,Schiminovich2001}. One of the few surveys at low Galactic latitudes was the spectral imaging survey of the {\it SPEAR} ({\it Spectroscopy of Plasma Evolution from Astrophysical Radiation}: \citet{Edelstein2006}) satellite at wavelengths between 900 and 1750 \AA. \citet{Park2012} combined {\it SPEAR} and {\it Galaxy Evolution Explorer} (\galex) data to study the diffuse UV radiation over a 30 square degree region in the vicinity of the Aquila Rift. They deduced that most of the continuum emission was due to dust-scattered light but with contributions from molecular hydrogen fluorescence over part of the field. We have focused on a part of this field roughly corresponding to the region they call ``Ophiuchus'' and have used a Monte Carlo model of dust scattering to model the emission observed in the far ultraviolet (FUV) and the near ultraviolet (NUV). \galex\ observed about 80\% of the sky before it was turned off on June 28, 2013 \citep{Murthy2014b} but had very few observations at low latitudes, primarily because of fears that the bright diffuse emission might damage the detectors. This constraint was relaxed late in the mission but the FUV power supply had failed by then. This region is the only location that included both FUV and NUV observations at a latitude below 10 degrees.

\section{Observations and Data}
\begin{figure*}
\centering
\includegraphics[width=3.4in]{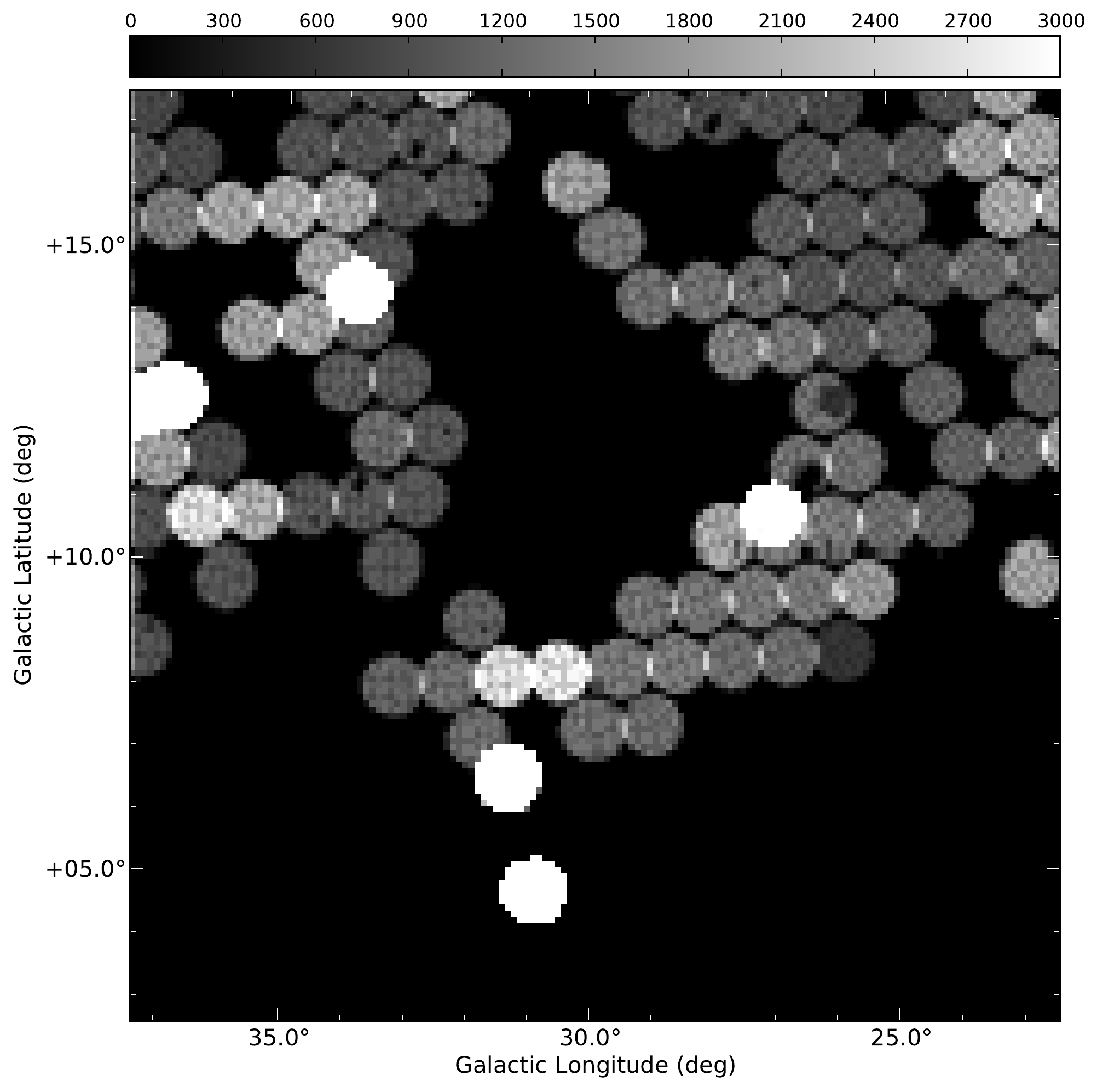}
\includegraphics[width=3.4in]{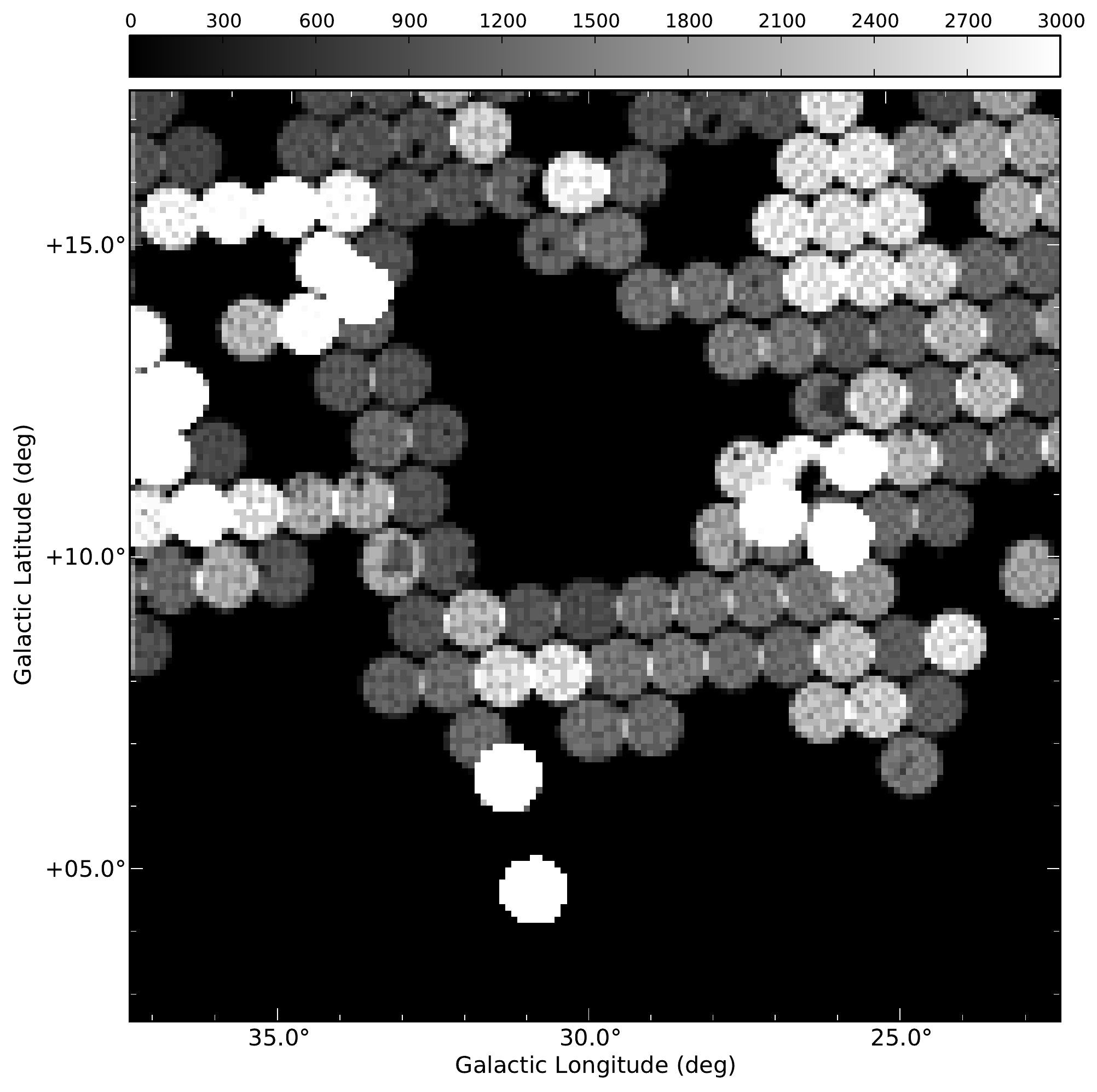}
\caption{FUV (left) and NUV (right) exposure time maps. The colour bars show the exposure times with black representing regions with no \galex\ coverage.}
\label{fig1}
\end{figure*}
\galex\ was a low Earth orbiting ultraviolet (UV) telescope under the NASA Small Explorer (SMEX) program whose primary goal was the observation of galaxies at low redshift. Along the way, it performed a survey of most of the sky in two UV passbands: the far ultraviolet (FUV: 1571 \AA) and the near ultraviolet (NUV: 2317 \AA). The spacecraft and the mission were described by   \citet{Martin2005} and the calibration and data products by \citet{Morrissey2007}. The final data release (GR6/GR7) included all observations before the completion of the mission in 2013, barring a few proprietary observations taken as part of the CAUSE (Complete All-Sky UV Survey Extension) program. \citet{Bianchi2014} and \citet{Seibert2012} have combined all the individual point sources detected in each observation into a single all-sky catalog and \citet{Murthy2014b} has produced a diffuse map of the entire sky in each of the two bands.

A \galex\ observation consists of one or more visits pointed toward a single location in the sky. Because the visits may be on different days, each visit will have  different contributions from the airglow and zodiacal light \citep{Murthy2014a} and so has to be processed independently if we want to extract the diffuse background. This was done by \citet{Murthy2014b} who ran all the GR6/GR7 visits through an automated pipeline. He blanked out point sources using the merged catalog for each observation, binned the pixels into 2$\arcmin$ pixels and subtracted the foreground airglow and zodiacal light using the formulation of \citet{Murthy2014a}. The resultant data product consists of a separate data file for each visit containing only the astrophysical background and is available from the Mikulski Archive for Space Telescopes (MAST)\footnote{https://archive.stsci.edu/prepds/uv-bkgd/}.

\begin{figure}
\centering
\includegraphics[width=3.4in]{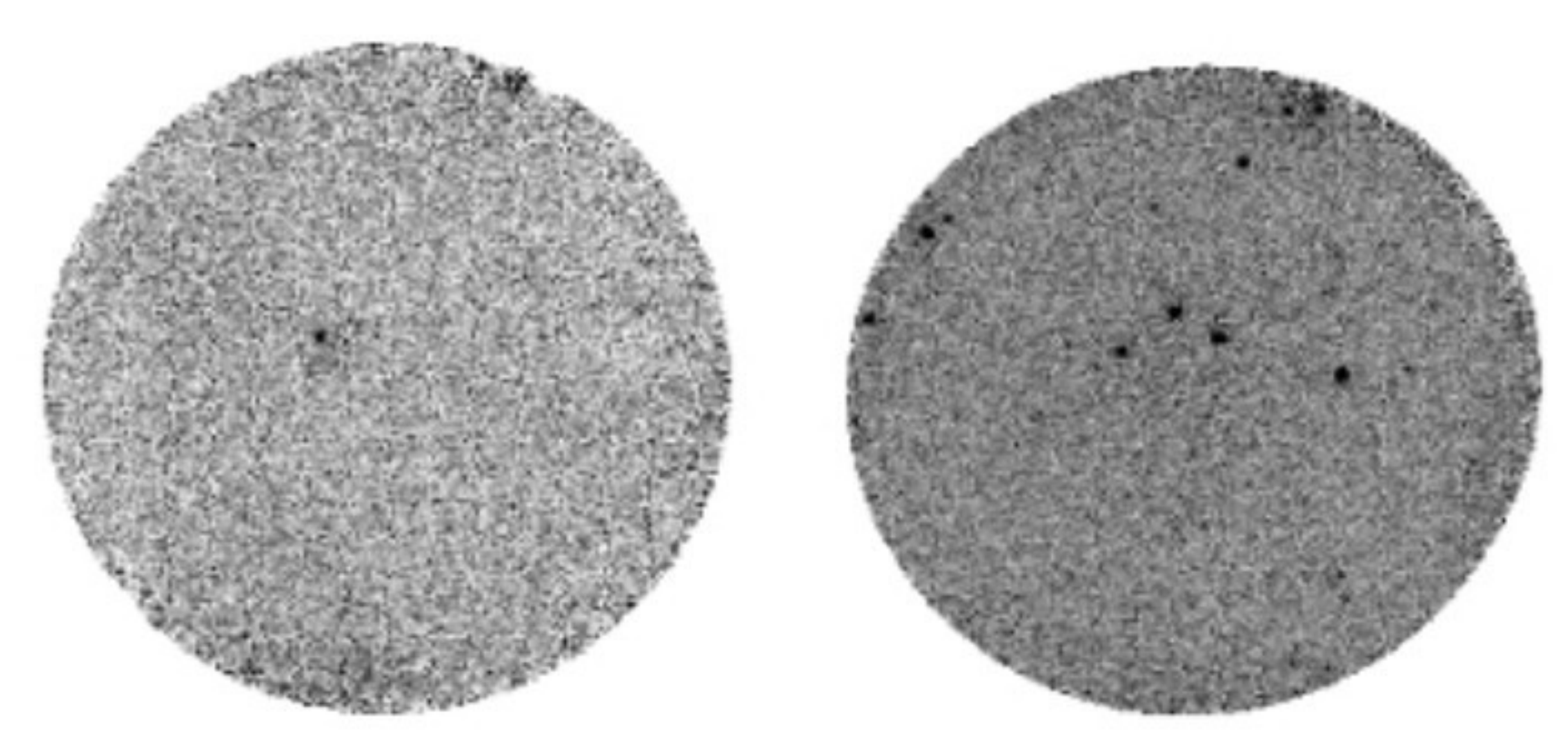}
\caption{Original \galex\ FUV image (left) and NUV image (right)  for the lowest latitude observation, COROT\_0507B.}
\label{fig2}
\end{figure}
\begin{figure*}
\centering
\includegraphics[width=3.4in]{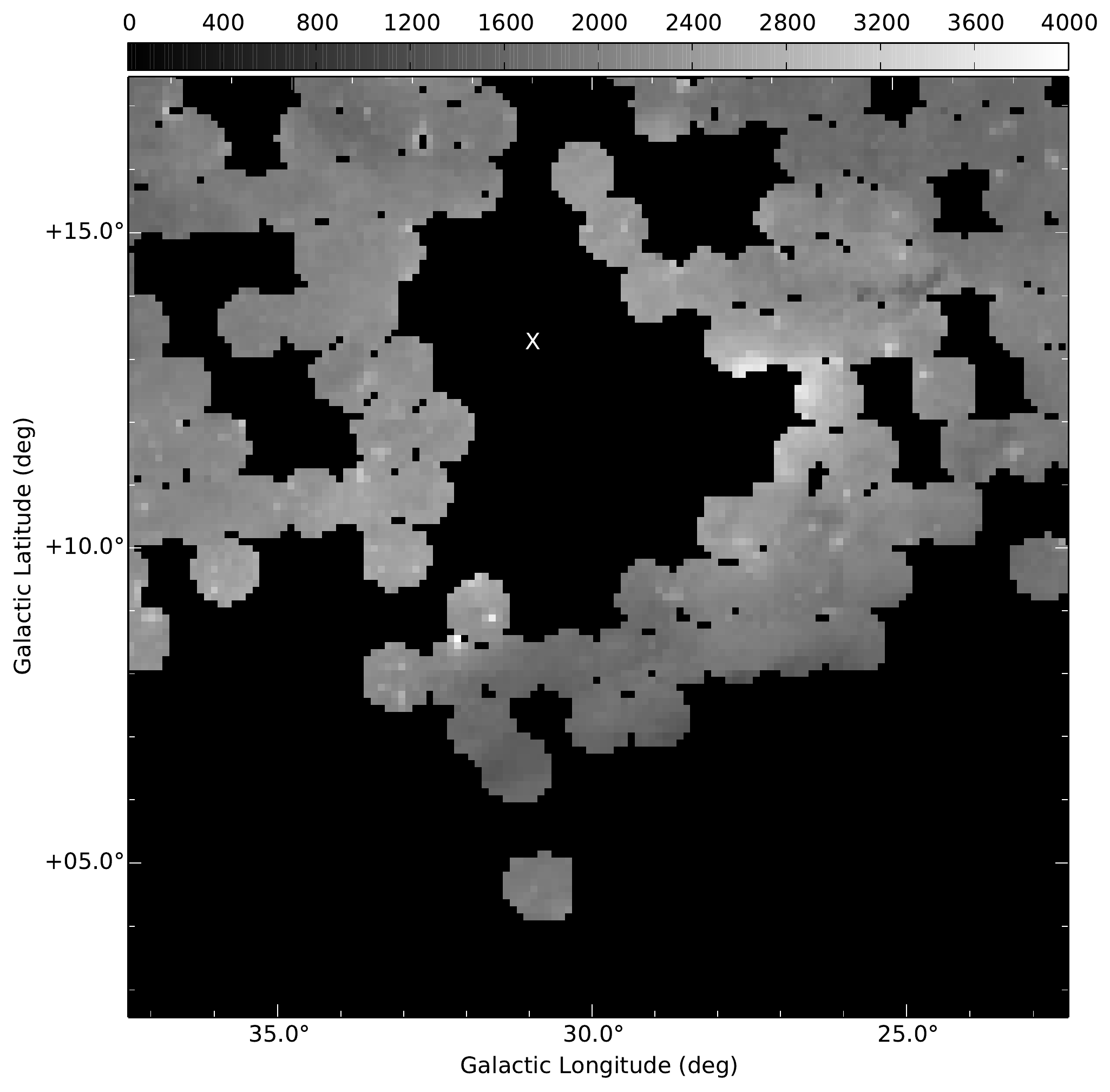}
\includegraphics[width=3.4in]{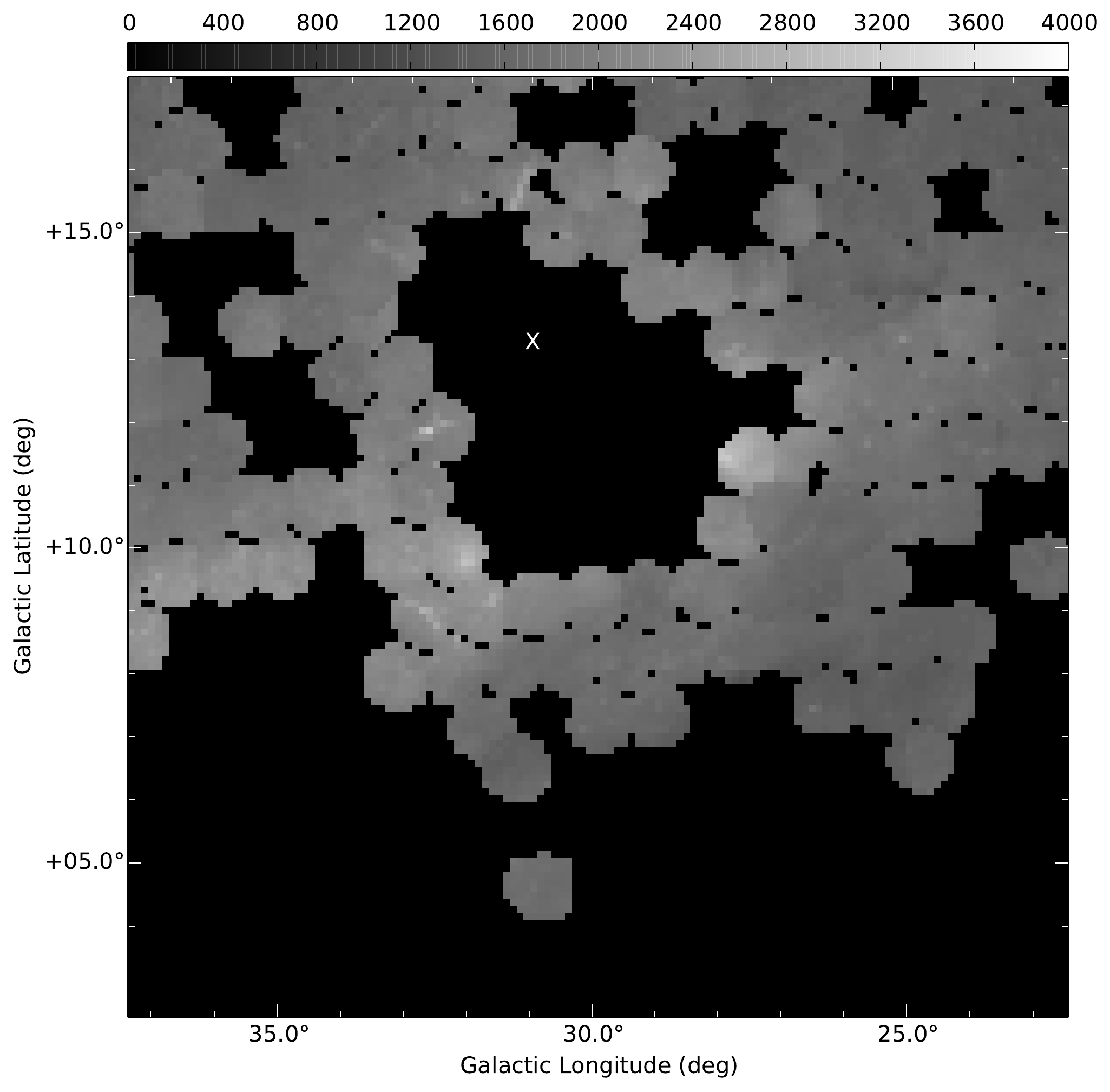}
\includegraphics[width=3.4in]{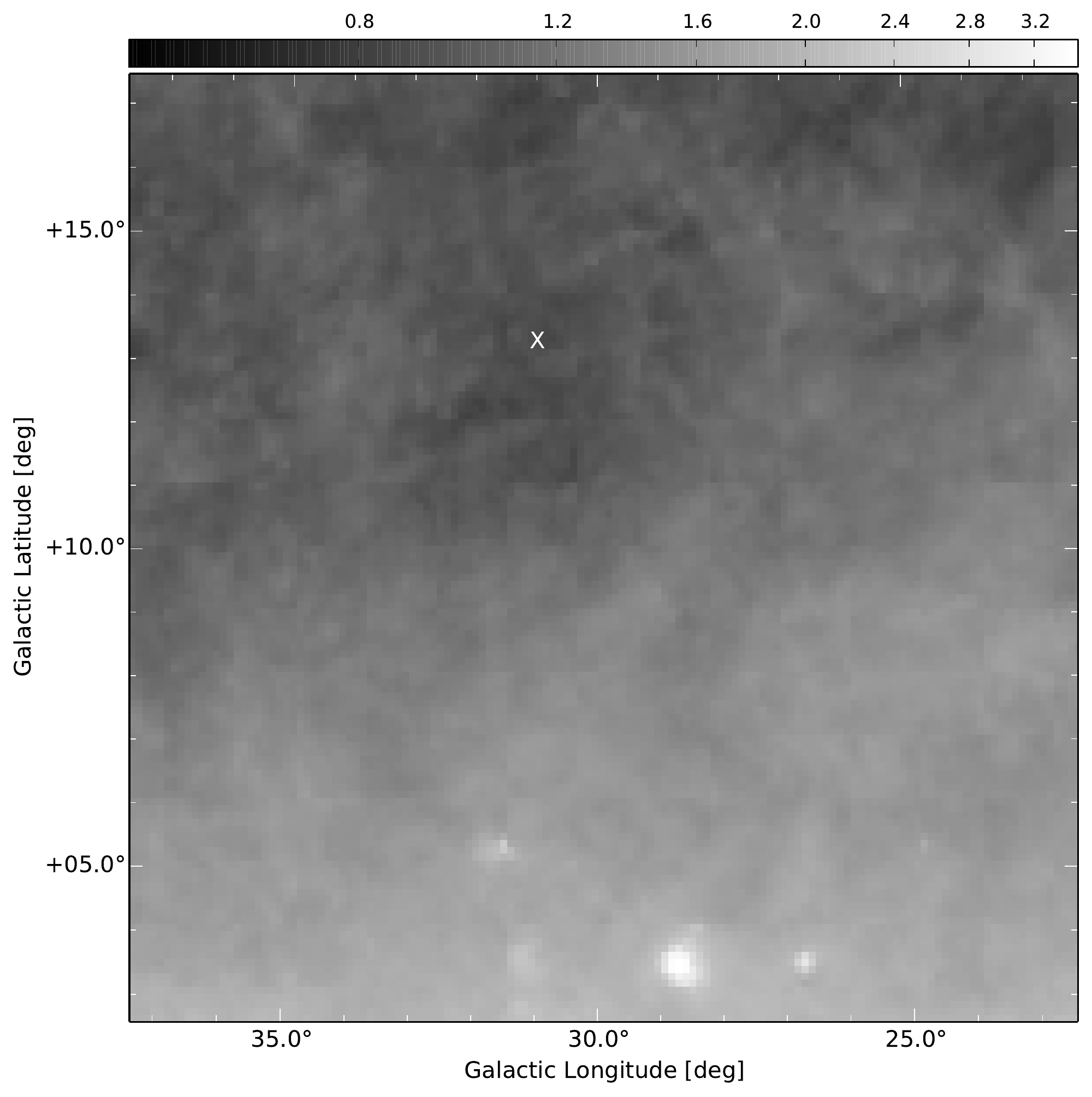}
\caption{FUV (top left) and  NUV (top right) diffuse maps of the 15$\degree$x15$\degree$ region centred at {\it l} = $30.0\degree$ and {\it b} = $10.0\degree$. The IRAS  $100 \micron$ map is at the bottom. There is no UV data in the centre of the region because of HIP 88149 (`X' in the map) and none at the lowest latitudes.}
\label{fig3}
\end{figure*}
This work comprises \galex\ observations of an area of 15 square degrees centred at {\it l} = $30.0\degree$ and {\it b} = $10.0\degree$. There were 227 observations with FUV data and 258 with NUV data, divided into 399 FUV visits and 567 NUV visits. Most of these observations were from the All-Sky Imaging Survey (AIS) with exposure times of 100 to 300 seconds but there were a few observations made by guest investigators with observation times of several thousand seconds each. The exposure time per pixel of the co-added data is shown in Fig.~\ref{fig1}. For reference, we have shown the lowest latitude observation, which is also the deepest, at the full \galex\ resolution in Fig. \ref{fig2}. Note that there are very few UV stars in the field, even at these low latitudes.

\begin{figure*}
\centering
\includegraphics[width=3.2in]{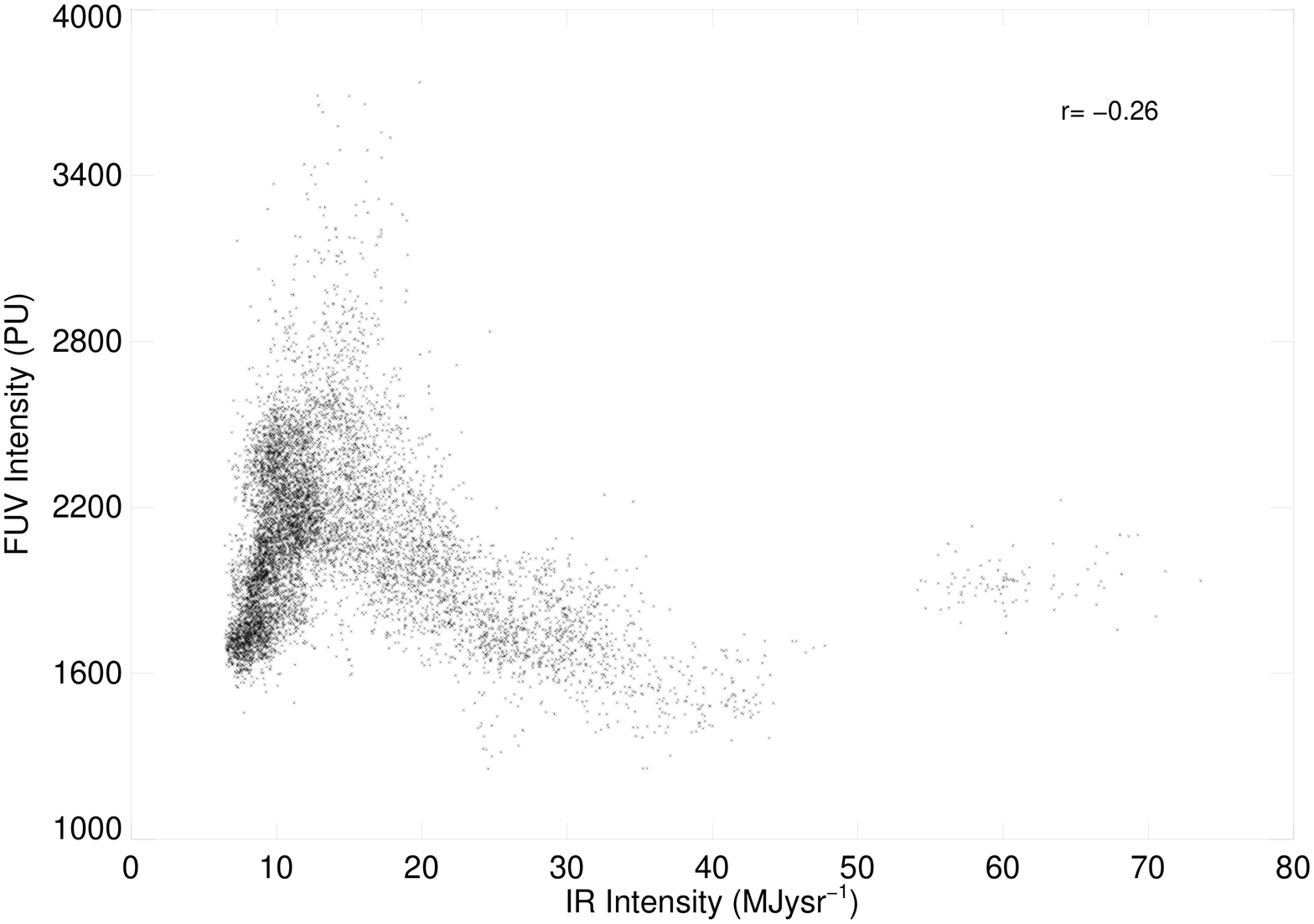}
\includegraphics[width=3.2in]{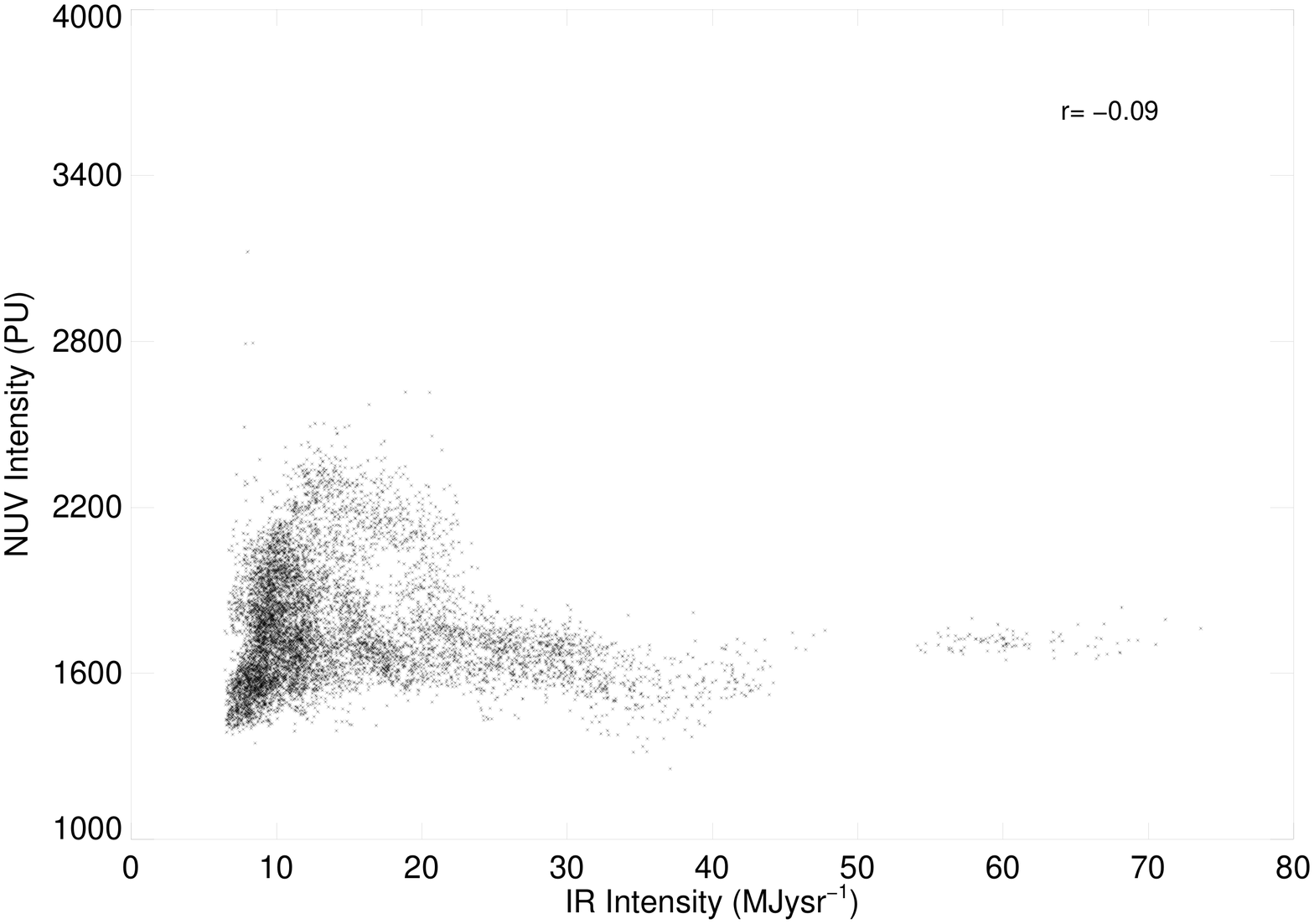}
\caption{GALEX FUV (left) and NUV (right) intensity plotted against IRAS $100 \micron$ intensity. The y axis is in photon units
(\photu).}
\label{fig4}
\end{figure*}

We binned the \galex\ data in the field into $6\arcmin$ bins to increase the signal-to-noise in each of the two UV bands weighting each pixel by its exposure time. We have only used the central $0.5\degree$ radius of each observation to avoid edge effects in the data, as recommended by \citet{Murthy2014b}. The resultant image in each of the two bands is shown in Fig.~\ref{fig3} along with the $100 \micron$ emission from the {\it Infrared Astronomy Satellite} ({\it IRAS}). There are two large exclusion zones in the UV data: in the centre around the star HIP 88149 (B2V, V$_{mag}$ = 4.6, d = 207 pc); and at the lowest latitudes where the IR flux rises rapidly. Despite these gaps, it is apparent that there is no correlation between the UV and the IR intensities (Fig. \ref{fig4}). This is not surprising as the cross-section of the grains is much less in the IR than in the UV \citep{Draine2003} implying that the thermal emission observed in the IR bands samples a much longer line of sight than the UV, which saturates over only a few hundred parsecs.

\section{Modelling and Results}

We have developed a model for the dust scattered radiation over the entire sky \citep{Murthy2015} and have used that model to predict the expected radiation in this region as a function of the Henyey-Greenstein \citep{Henyey1941} optical parameters: the albedo ({\it a}) and the phase function asymmetry factor ({\it g}). The albedo is the amount of light reflected by the grains while {\it g} ($= <cos\theta>$) governs the angular dependence of the scattered radiation. Isotropic scattering is represented by $g = 0$ with the amount of forward scattering increasing with increasing {\it g}. Although there are suggestions that the actual scattering is more complex \citep{Draine2003}, the data have not been of sufficient quality to justify using more complicated scattering functions.

\begin{figure}
\centering
\includegraphics[width=3.2in]{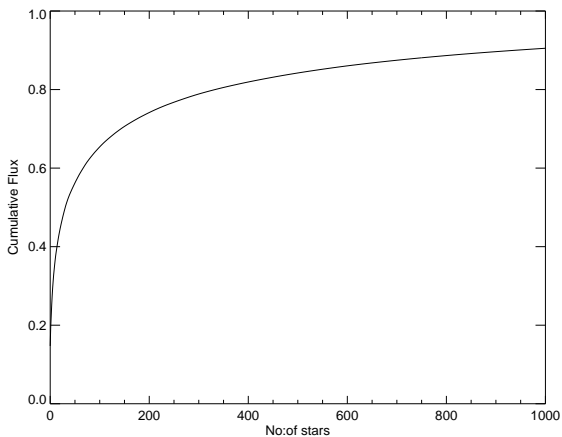}
\caption{The cumulative contribution to the flux from stars in the Galaxy  shown. }
\label{fig5}
\end{figure}

\begin{figure}
\centering
\includegraphics[width=3.2in]{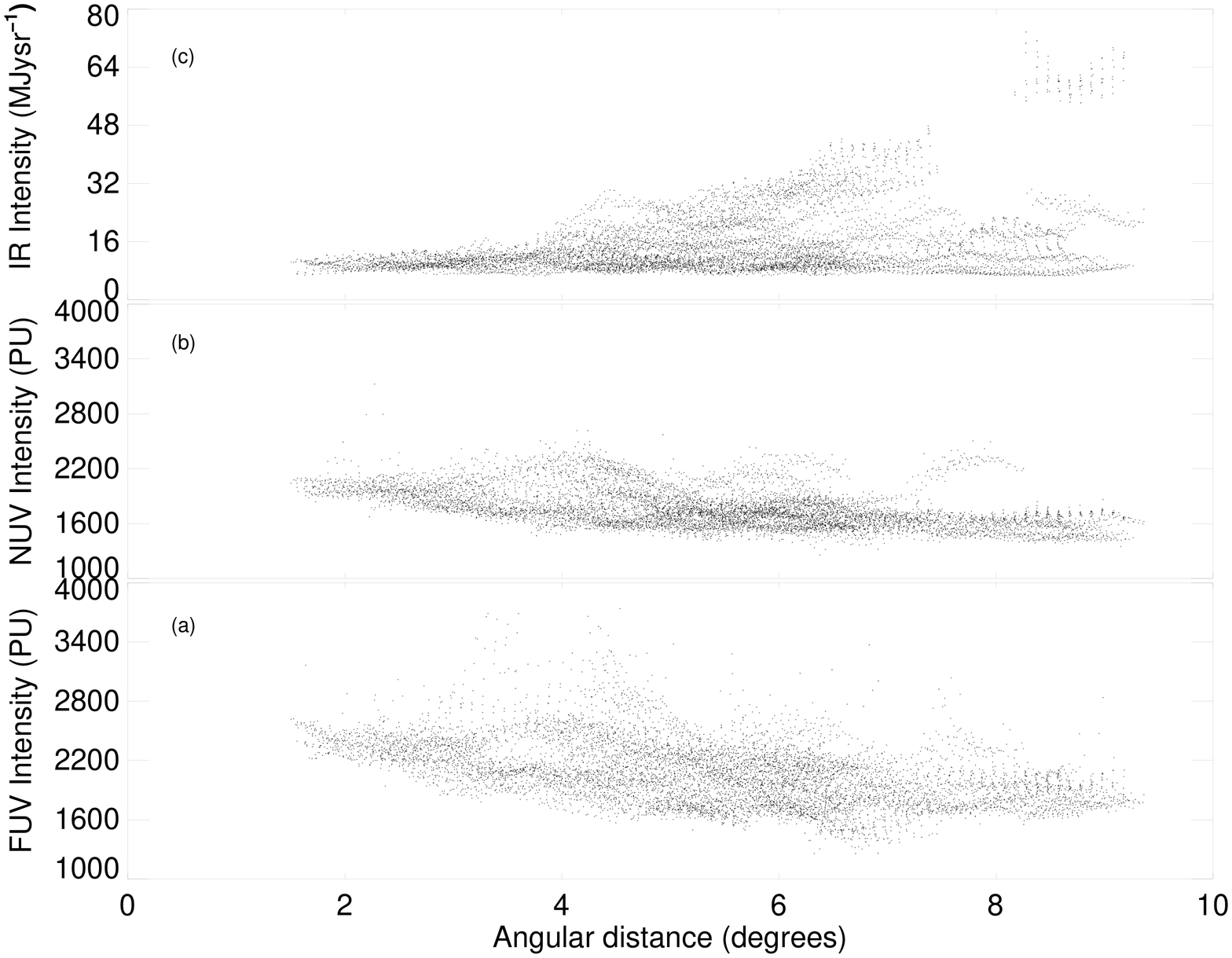}
\caption{ (a) \galex\ FUV (b) \galex\ NUV and (c) IRAS $100\micron$ intensities plotted against angular distance from HIP 88149. PU = \photu .}
\label{fig6}
\end{figure}

The photons contributing to the diffuse background come from a relatively small number of O and B stars \citep{Murthy1995} for which a catalogue integration works well \citep{Henry1977}. Here, we have used stars from the Hipparcos catalogue \citep{Perryman1997} whose spectral type, distance, and B and V magnitudes are tabulated. The number of photons from each star at the desired wavelength were calculated using model spectra from \citet{Castelli2004}, as described by \citet{Sujatha2004}. Although stars from the entire Galaxy do contribute to the diffuse light, the diffuse flux in this field is dominated by a much smaller number of stars (Fig. \ref{fig5}) with 35\%\ of the flux originating from just 10 stars and 65\%\ from 100 stars. The brightest of these is HIP 88149 (66 Oph) which contributes 14\% of the total diffuse flux in the region and is responsible for the large exclusion zone in the centre of the UV observations. It is apparent from Fig. \ref{fig3} and Fig. \ref{fig6} that there is an enhancement of the diffuse flux in the UV --- but no significant enhancement in the IR --- around this exclusion zone.

\begin{figure}
\centering
\includegraphics[width=3.2in]{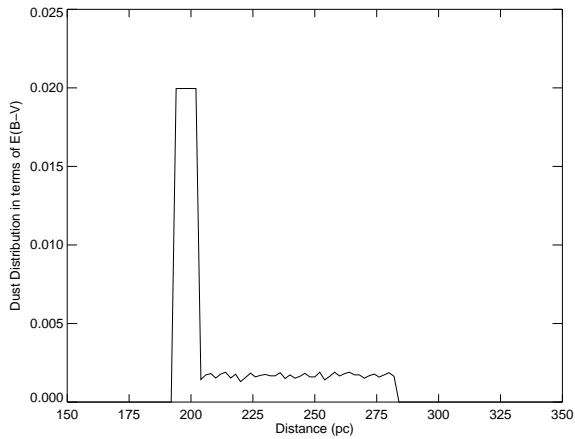}
\caption{Dust distribution as a function of distance to the cloud in a single line of sight.}
\label{fig7}
\end{figure}

\begin{figure}
\centering
\includegraphics[width=3.2in]{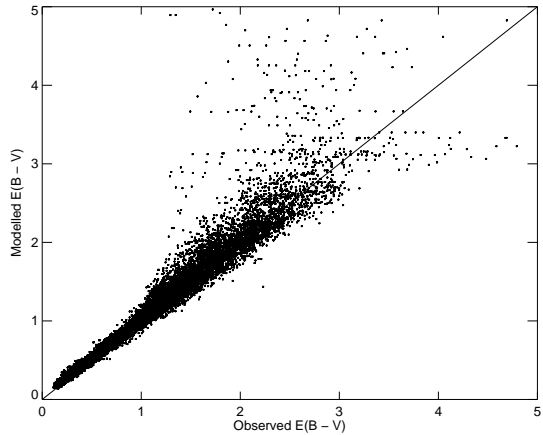}
\caption{Comparison of the modelled E(B - V) with the observed E(B - V) from Schlegel.}
\label{fig8}
\end{figure}

For the sake of this work, we have divided the sky into a 500x500x500 grid with a bin size of 2 pc on each side and filled each pixel with a hydrogen density corresponding to density of 1 cm$^{-3}$ in the Galactic Plane with a scale height of 120 pc \citep{Lockman1984} and a cavity of 50 pc in radius around the Sun \citep{Welsh2010}. We convert this into an optical depth for the dust using the ``Milky Way'' model of \citet{Draine2003}, which implicitly assumes a uniform gas-to-dust ratio (N(H)/E(B-V)) of $5.8 \times\ 10^{21}$ \citep{Bohlin1978} throughout the Galaxy. More important to the scattered light in the observed region is the local distribution of dust as a function of distance. \citet{Straizys2003} placed the bulk of the material  in a cloud with its near edge at a distance of $225 \pm 55$ pc away from the Sun and a thickness of 80 pc. However, this would have placed most of the dust behind HIP 88149 which is at a distance of 207 pc from the Hipparcos catalogue \citep{Perryman1997} with an E(B - V) of 0.19 magnitudes \citep{Gudennavar2012}. We have therefore divided the dust into two clouds with a total reddening in each pixel given by the all-sky map of \citet{Schlegel1998}. We placed the first cloud at a distance of 195 pc from the Sun with a thickness of 10 pc and a maximum reddening of 0.2 magnitudes. The remaining dust, if any, was placed in the second cloud with a near edge of 205 pc and a thickness of 80 pc (Fig. \ref{fig7}). The spatial distribution of the modelled E(B - V) is in good agreement with the observed reddening (Fig. \ref{fig8}).

\begin{figure}
\centering
\includegraphics[width=3.2in]{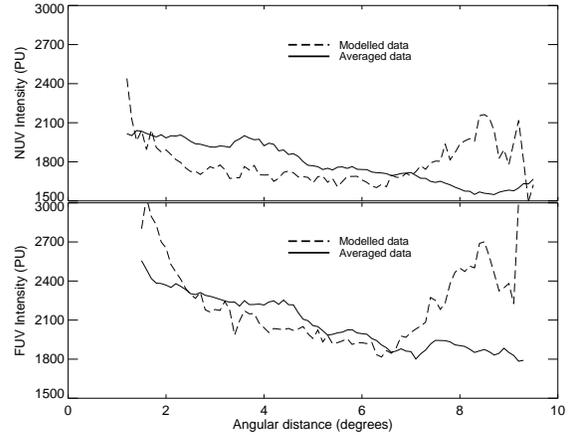}
\caption{Modelled FUV and NUV emission as a function of angular distance from HIP 88149 overplotted with the original FUV and NUV emission where the y-axis is in \photu .}
\label{fig9}
\end{figure}

\begin{figure}
\centering
\includegraphics[width=3.2in]{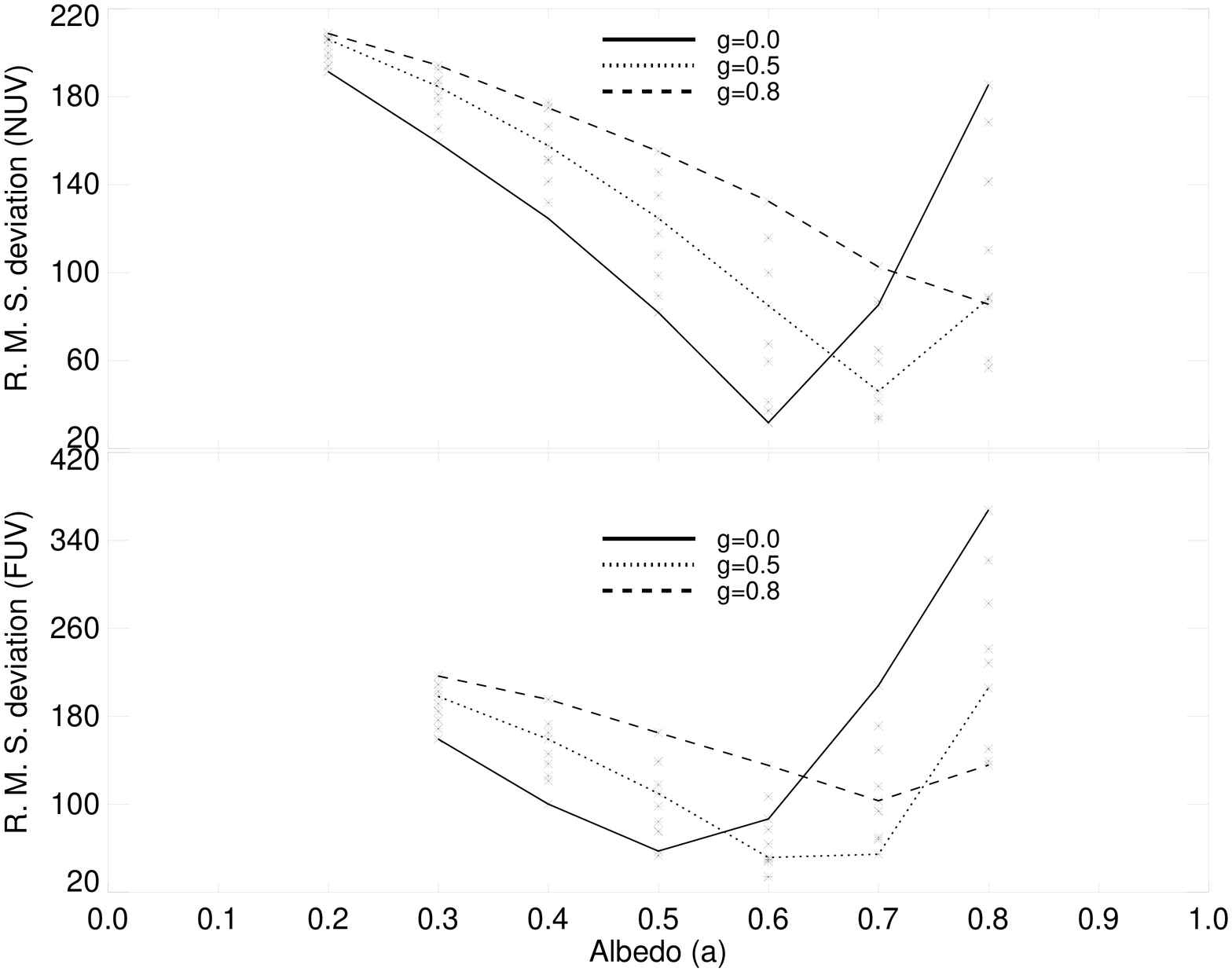}
\caption{Root mean square deviations as a function of albedo ({\it a}) for different {\it g} values. Lines
connect points with identical $g$.}
\label{fig10}
\end{figure}

\begin{figure}
\centering
\includegraphics[width=3.2in]{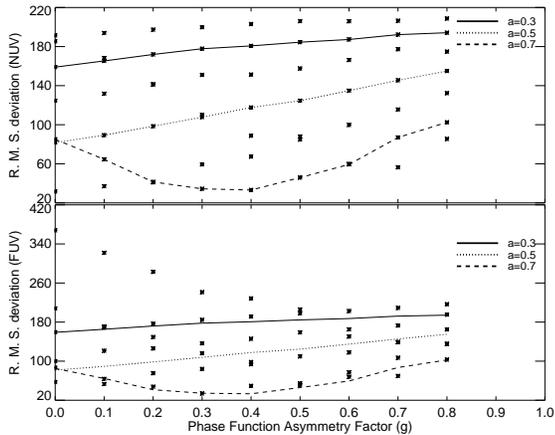}
\caption{Root mean square deviations as a function of phase function asymmetry factor ({\it g}) for different {\it a} values. 
Lines connect points with identical $a$.}
\label{fig11} 
\end{figure}

We have run a series of simulations with different values of the optical constants to model the emission in the Aquila Rift. Because of time limitations, each run was for 50 million photons over the entire grid out of which only about 0.02\% contributed to the flux in our field. At this stage in our modelling, it is difficult to match the data on a pixel by pixel scale and we have instead used the flux as a function of distance from HIP 88149 as a measure of the quality of the fit (Fig. \ref{fig9}). We have plotted the root mean square of the deviations of the model from the observed data as a function of albedo ($a$) in Fig. \ref{fig10} and as a function of $g$ in Fig. \ref{fig11} for both the FUV and the NUV channels. The best fit model data implies an albedo ($a$) of 0.6 - 0.7 for both channels and a phase function asymmetry factor ($g$) of 0.2 -- 0.4.

\begin{figure}
\centering
\includegraphics[width=2.9in]{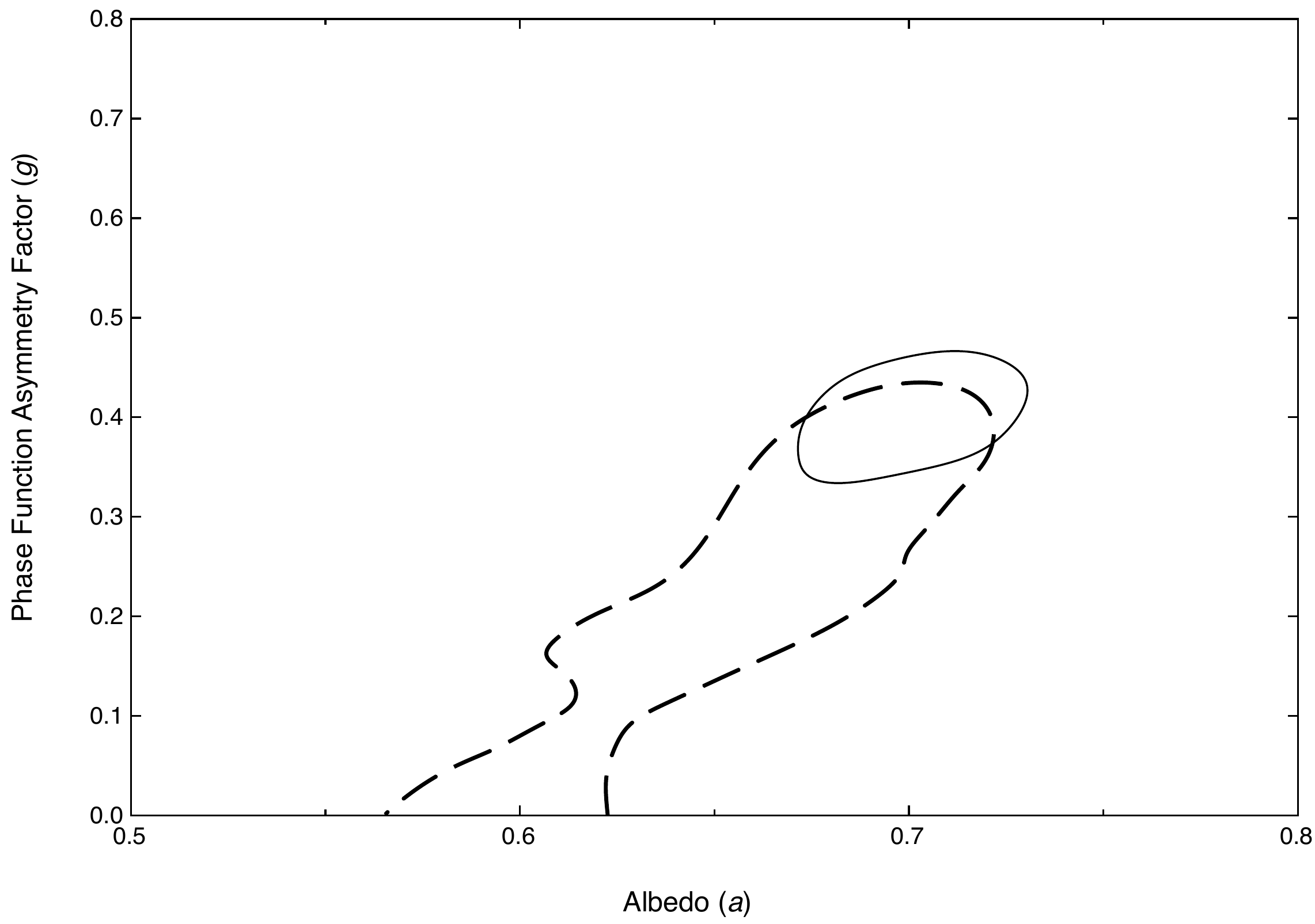}
\caption{Estimated $3\sigma$ confidence contours for $a$ and $g$ for FUV (solid line) and NUV (dashed line).}
\label{fig12}
\end{figure}

\begin{figure}
\centering
\includegraphics[width=3.2in]{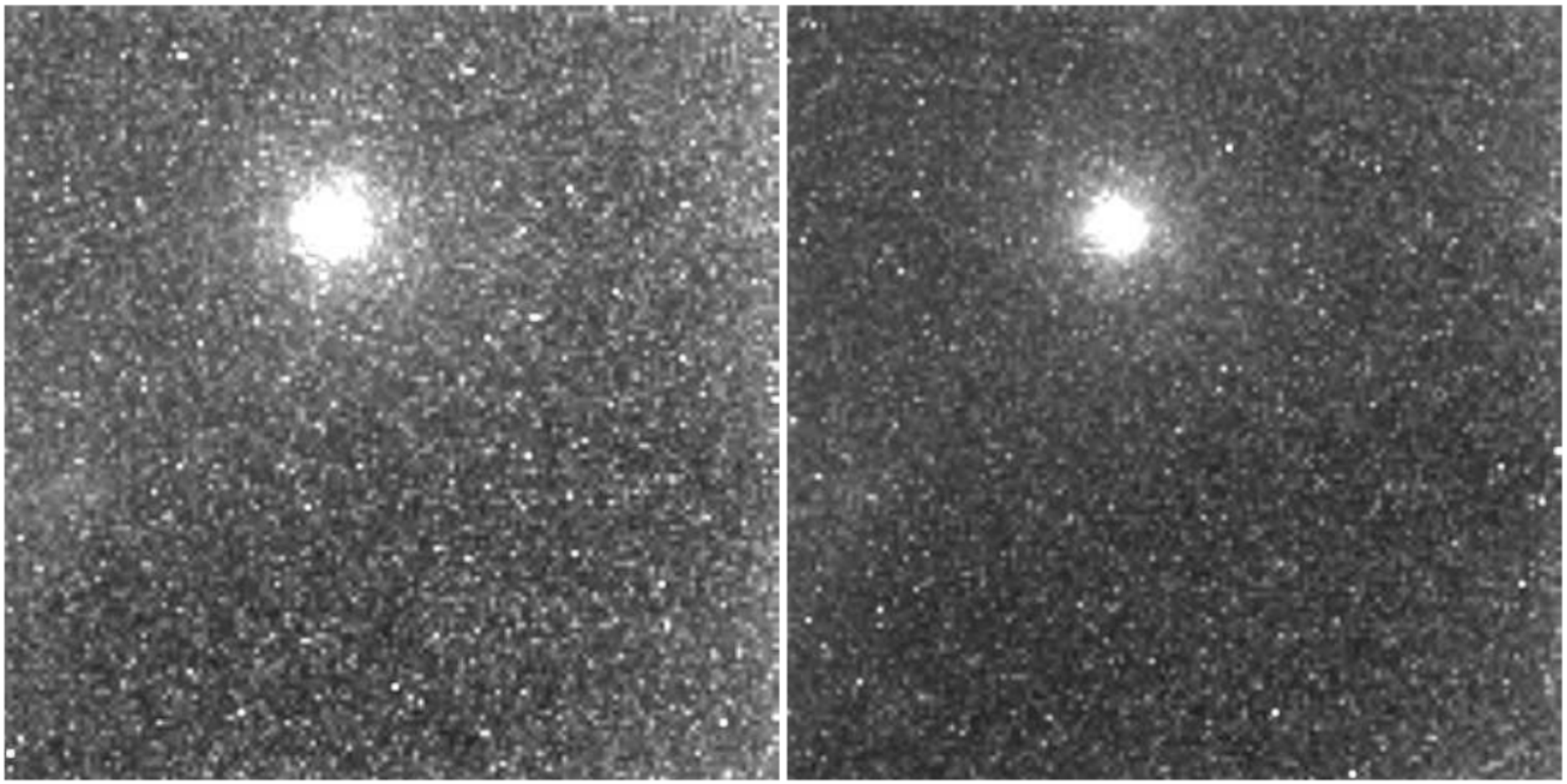}
\caption{Modelled FUV image (left)  and NUV image (right).}
\label{fig13}
\end{figure}

Our current models are inefficient in that most of the computational time is spent in modelling the entire Galaxy and not just the Aquila Rift region.  We have run much longer simulations ($> 10^{9}$ photons) for a more restricted set of optical constants and these are instructive in understanding the nature of the diffuse radiation. In each case, there is a large, bright halo around HIP 88149 (Fig. \ref{fig13}) the edges of which we see in the data. There is a trade-off in the modelled data between the albedo and $g$ in that the level of the diffuse radiation increases with albedo, as expected, but decreases with increasing $g$ due to the sharpening of the halo. This is illustrated in the $3 \sigma$ contours for each of the FUV and the NUV in Fig. \ref{fig12}, estimated using the procedure of \citet{Lampton1976}. These limits do not take into account the uncertainty in the dust distribution.  Although the bright scattering nebula around HIP 88149 is hidden in the exclusion zone of \galex, because of the safety restrictions on observing near bright stars, it is prominent in the {\it SPEAR} data of \citet{Park2012} in the region they call ``Ophiuchus''. Note that there is no dependence of the modelled or the observed flux on the Galactic latitude.


\begin{table}
\caption{Previously determined {\it a} and {\it g} values}
\label{ag}
\begin{tabular}{c c c}
\hline
{\it a} & {\it g} & Reference\\
\hline
$0.6 - 0.7$ & $ 0.2 - 0.4$ & This paper\\
$0.28 \pm 0.04$ & $0.61 \pm 0.07$ & \citet{Sujatha2007}\\
$0.40 \pm 0.10$ & $0.55 \pm 0.25$ & \citet{Sujatha2005}\\
$0.40 \pm 0.20$ & -- &\citet{Shalima2004} \\
$0.45 \pm 0.05$ & $0.68 \pm 0.10$ &\citet{Witt1997} \\
$0.45 \pm 0.15$ & -- & \citet{Murthy1995}\\
$0.50$ & $0.90$ &\citet{Witt1994} \\
$0.47-0.70 (136.2  nm)$ & $ $ & \\
$0.55-0.72 (176.9  nm)$ & $0.0-0.80$ &\citet{Gordon1994} \\
$0.42$ & $0.44$ & \citet{Wright1992}\\
$0.50$ & $0.50$ &\citet{Morgan1978} \\
$0.60 \pm 0.05$ & $0.75 \pm 0.15$ & \citet{lillie1976}\\
\hline
\end{tabular}  
\end{table}

There are several uncertainties in our modelling, the most important of which is that we do not know the distribution of the dust in any given line of sight \citep{Seon2011}. Our modelling suggests that we require a substantial amount of dust near HIP 88149 in order to reproduce the observed halo around the star, yet we do not see a corresponding infrared halo which would imply that the dust cannot be too near the star. As mentioned above, our current simulation wastes most of the photons in distant parts of the sky but because more than 10\% of the diffuse flux in the region is due to multiply scattered photons from outside the field of view, it is not trivial to reduce the scope of the simulation. This points to the immediate improvements to be made in our model. The first, as noted above, is to make the program more efficient by restricting the modelling to a much smaller area taking into account the fact that many of the scattered photons are non-local. The second is to improve the dust distribution. One important new input is the all-sky dust map based on PAN-STARRS and 2MASS photometry by \citet{Green2015} which in combination with the limits put by the UV scattering observations may help to pin down the dust distribution and properties in this region.

\section{Conclusions}
We have used archival \galex\ data combined with a Monte Carlo dust scattering model to probe the diffuse radiation in one field at low Galactic latitudes ({\it b} $< 15\degree$). Although the infrared 100 \micron\ flux falls off rapidly from the Galactic Plane, there is no corresponding dependence of the UV flux on the latitude, at least in this restricted region near the Aquila Rift. Approximately 15\% of the flux comes from HIP 88149, a B2 star in the field of view, which is responsible for an extended reflection nebula visible to a distance of more than 2$\degree$ from the star. We fit the overall level of the diffuse background with an albedo of between 0.6 -- 0.7 with a $g$ between 0.2 -- 0.4. The albedo is very dependent on the dust distribution in the line of sight and is therefore correspondingly uncertain. However, the value of $g$ is constrained by the extent of the halo which would be too small if $g$ were larger. These values are in reasonable agreement with other determinations of the albedo in a variety of regions (Table \ref{ag}).

Although the UV background follows a superficial correlation with the thermal IR emission over the entire Galaxy \citep{Hamden2013, Murthy2014b}, there are clearly wide variations in the ratio. \citet{Henry2015} suggested that these were due to a different, unknown contributor to the diffuse UV which was associated with the Galaxy but not part of the dust which contributes to the IR emission. They also pointed out that separate analyses of the background in small areas might individually be able to fit the flux by dust scattering but may not be consistent across regions.

We find here that single stars may be the prime contributors to the diffuse background over a wide area. Although at low Galactic latitudes, the scattered light is from a dust cloud at a distance of 200 pc from the Sun while the bulk of the IR emission is from an integrated column through much longer distances. Thus a complete understanding of the UV scattering depends on modelling both the diffuse emission over the entire sky as well as in individual locations where local conditions are important. This has important implications in the planning of observations with future UV missions such as the upcoming ASTROSAT mission \citep{Kumar2012}. The Galactic plane is excluded on safety grounds because of the presumed brightness of the UV background, as was done with \galex. Our work here and with {\it Voyager} \citep{Murthy1999} shows that there are dark regions even in the Galactic Plane.

\section*{Acknowledgement}

This work is based on the data from NASA's \galex\ spacecraft. \galex\ is operated for NASA by the California Institute of Technology under NASA contract NAS5-98034. We acknowledge the NASA's Astrophysics Data System and NASA's SkyView facility (http://skyview.gsfc.nasa.gov) located at NASA Goddard Space Flight Center. Jyothy is grateful to Sri Mata Amritanandamayi Devi, the Chancellor of Amrita University for her spiritual guidance. Jyothy would like to thank Mr. Sathyanarayanan for his help. We thank the referee for a prompt response and for pointing out inconsistencies in the first draft. 
\nocite{*}
\bibliographystyle{mn2e}
\bibliography{references}
\label{lastpage}

\end{document}